\documentclass[aps,prb,twocolumn,groupedaddress]{revtex4-1}  
\usepackage{graphicx}  
\usepackage{epstopdf}
\usepackage[caption=false]{subfig}
\usepackage{float}
\usepackage[toc,page]{appendix}
\usepackage{cleveref}
\usepackage{dcolumn}   
\usepackage{bm}        
\usepackage{amssymb}   
\usepackage{mathrsfs}   
\usepackage{color}
\usepackage{hyperref}
\usepackage{amsmath, amsthm, amssymb}
\usepackage{amsfonts}
\usepackage{bbold}       
\usepackage{accents}    
\newcommand{\ut}{\underaccent{\tilde}}

\usepackage{soul}
\newcommand{\beq}{\begin{equation}}
\newcommand{\eeq}{\end{equation}}
\newcommand{\bqa}{\begin{eqnarray}}
\newcommand{\eqa}{\end{eqnarray}}
\newcommand{\nn}{\nonumber}

\newcommand{\erf}[1]{Eq.~(\ref{#1})}

\newcommand{\erfa}[2]{Eqs.~(\ref{#1}) and (\ref{#2})}
\newcommand{\arf}[1]{Appendix~\ref{#1}} 
\newcommand{\srf}[1]{Sec.~\ref{#1}} 
\newcommand{\crf}[1]{Ref.~\cite{#1}} 
\newcommand{\frfs}[2]{Figs.~(\ref{#1})--(\ref{#2})}
\newcommand{\frf}[1]{Fig.~\ref{#1}}

\newcommand{\dg}{^\dagger}

\definecolor{BLACK}{gray}{0}
\definecolor{RED}{rgb}{1,0,0}
\definecolor{GREEN}{rgb}{0.2,.6,0.2}
\definecolor{BLUE}{rgb}{0,0,1}

\newcommand{\blk}{\color{BLACK}}

\newcommand{\sq}[1]{\left[ {#1} \right]}

\newcommand{\ro}[1]{\left( {#1} \right)}
\newcommand{\an}[1]{\left\langle{#1}\right\rangle}

\newcommand{\bra}[1]{\langle{#1}|}
\newcommand{\ket}[1]{|{#1}\rangle}
\newcommand{\ip}[2]{\langle{#1}|{#2}\rangle} 
\newcommand{\op}[2]{\ket{#1}\bra{#2}}
\newcommand{\s}[1]{\hat \sigma_{#1}}

\newcommand{\tp}{^{\top}} 

\begin{document}

\widetext


\title{Stochastic feedback control of quantum transport to realize a dynamical  ensemble 
of two nonorthogonal pure states}
\author{Shakib Daryanoosh} \email{shakib.daryanoosh@griffithuni.edu.au}
\author{Howard M.~Wiseman} \email{H.Wiseman@griffith.edu.au}
 \affiliation{Centre for Quantum Computation and Communication Technology (Australian Research Council),Centre for Quantum Dynamics, Griffith University, Brisbane, Queensland 4111, Australia}
 \author{T.~Brandes} \email{brandes@physik.tu-berlin.de}
 \affiliation{Institut f\"ur Theoretische Physik, Hardenbergstr. 36, TU Berlin, D-10623 Berlin, Germany}
\date{\today}

\begin{abstract} 
	A Markovian open quantum system which relaxes to a unique steady state $\rho_{\rm ss}$ of finite rank  
	can be decomposed into a finite physically realizable ensemble (PRE) of pure states. That is, 
	as shown by Karasik and Wiseman [\href{http://prl.aps.org/abstract/PRL/v106/i2/e020406}
	{Phys. Rev. Lett. {\bf 106}, 020406 (2011)}], in principle there is a way 
	to monitor the environment so that in the long time limit the conditional state jumps between a finite number of 
	possible pure states.  In this paper we show how to apply this idea 
	to the dynamics of a double quantum dot arising from the feedback control of quantum transport,  
	as previously considered by one of us and co-workers [\href{http://prb.aps.org/abstract/PRB/v84/i8/e085302}
	{Phys. Rev. B {\bf 84}, 085302 (2011)}]. 
	Specifically, we consider the limit where the system can be described as a qubit, and show that while the 
	control scheme can always realize a two-state PRE, in the incoherent tunneling regime there are infinitely many PREs 
	compatible with the dynamics that cannot be so realized. For the two-state PREs that are realized, 
	we calculate the counting statistics and see a clear distinction between the coherent and incoherent regimes. 
\end{abstract}

\maketitle

\section{Introduction}
An open quantum system  undergoes a continuous
interaction with its environment and typically becomes entangled with it. \blk 
Although the state of total system---composed of physical system and 
its bath---{may be} pure, the state of system itself {becomes} mixed. 
In this paper we restrict to Markovian systems, by which we mean ones where 
the interaction with the environment is well modelled by quantum white noise 
~\cite{GarZol00,WisMil10}. \blk 
{In this model, the system continuously leaks quantum } \blk
information into the bath,  which is why the system state loses purity. This loss of purity can, 
however, be \blk compensated for
by monitoring the environment, over time scales sufficiently short in comparison to  the \blk system
evolution time, from which one gains information about the system. In this process, assuming
perfect detection, the state of system continuously collapses into pure states $\rho_c$ conditioned
on measurement results so that on average the unconditional  mixed \blk state $\rho$ is 
retrieved \cite{WisMil10}. This  so-called quantum trajectory theory~\cite{Car93}, \blk 
{describes both quantum jumps}~ \cite{DalCasMol92,GarParZol92,Bar93}, 
and quantum diffusion~\cite{Car93,GisPer92b,WisMil93c}. These are examples of different 
types of ``unravelling''~\cite{Car93} of the master equation describing the evolution of $\rho$. \blk 

In this work we are concerned with Markovian open quantum systems  that are 
finite-dimensional  and have ergodic evolution. \blk Any such system 
ultimately relaxes to its unique steady state $\rho_{ \rm ss}$ which
can be expressed {as a} \blk statistical mixture of a  finite ensemble of pure states. 
{For any given $\rho_{ \rm ss}$, \blk
	there are  an \blk infinite number of such decompositions, 
	because the states therein may be nonorthogonal. 
	However, there is a crucial distinction to be made here: only some of these 
	ensembles will be {\em physically realizable}~\cite{WisVac01} as the possible conditioned states 
	of the system under some unravelling. \blk It was shown how to construct {\em physically realizable ensembles} (PREs), 
	but the question of {\em how} to physically realize them (i.e. what type of unravelling) was not addressed in general. 
	For finite dimensional systems it was subsequently shown~\cite{KarWis11} that achieving a particular finite PRE will, 
	in general, require an adaptive monitoring. 
	
	The concept of an adaptive measurement is a form of quantum control that is related to, but distinct from, 
	feedback control~\cite{WisMil10}.  \blk 
	Feedback control  to change the dynamics of open   quantum systems is an ongoing
	research topic of with applications in noise reduction, state purification, and state protection~\cite{WisMil10,Wis94,Hor97,Com10,PolEmaBra11,Gil10,Say11}.  
	Adaptive measurements, by contrast, do not change the dynamics but rather change the 
	quality of the information obtained about the system, with application in metrology~\cite{Wis95,Bra07} 
	and measurement-based quantum computation~\cite{RauBri01,PreZei07}. \blk 
	
	The central question to be examined in this study is whether  the introduction of feedback control in \blk 
	a Markovian open quantum system  can result in conditional dynamics that 
	realize a  PRE that is finite, and is nonclassical, in the sense of comprising 
	nonorthogonal states. \blk We address this by organising the structure of the paper 
	as follows. First, \srf{sec:II} reviews how it is possible to track a finite { dimensional} open quantum system 
	{to produce a} \blk finite PRE.  In particular, we reproduce the result~\cite{KarWis11} 
	that for a qubit there always exists a two-element PRE; that is, a bit is sufficient to  
	keep track of a qubit's \blk dynamics . Next, in \srf{sec:III} we present the proposed idea of 
	implementing feedback control on quantum transport in mesoscopic systems to purify the state of the system 
	\cite{PolEmaBra11}. These two {are united in} \blk \srf{sec:IV} 
	where we show that the conditional dynamics for this mesoscopic system, with feedback, 
	can be used to realize some of the PREs that are theoretically possible, but not all of them. 
	\srf{sec:V} discusses the experimental signature of our scheme: the cumulants for 
	the counting statistics of quantum transport. \blk Finally, the conclusions are presented. 
	\section{Tracking the state of a qubit using a bit} \label{sec:II}
	In this paper we consider Markovian open quantum systems,  so that the state
	of system is determined, in  a suitable interaction frame, by  a Linbladian master 
	equation  
	\beq \label{eqn:MEL}
	\dot{\rho} = -i [\hat H,\rho] + \sum_{l} {\cal D} [\hat{c}_{l}] \rho \equiv {\cal L}\rho, \blk
	\eeq
	where ${\cal D}[\hat{O}] \bullet = \hat{O} \bullet {\hat{O}}\dg - 1/2\{\hat{O}\dg \hat{O},\bullet \}$ is the usual irreversible superoperator~\cite{WisMil10}. In the above, 
	 $\hat H$ is the system Hamiltonian,  
	and the collection of Lindblad operators $\{\hat{c}_l\}$ is determined by the coupling of the system to the bath. The Lindbladian superoperator 
	${\cal L}$ can also be expressed as 
	\beq   \label{eqn:ME}
	{\cal L} =  {\cal S}[\hat H'] + \sum_{l=1}^L {\cal J}[\hat{c}_l]
	\eeq \blk 
	where ${\cal S}[\hat H']\bullet = (- i  \hat{H}^\prime) \bullet + \bullet (i\hat{H}^\prime{}\dg )$ and 	${\cal J}[\hat{c}]\bullet = \hat{c} \bullet \blk\hat{c}\dg$  are superoperators which describe smooth evolution and 
	dynamical jumps, respectively (see below). Here 
	\beq \label{nonHH}
	\hat{H}^\prime = \hat{H} - \frac{i}{2} \sum_{l=1}^L \hat{c}_l\dg \hat{c}_l
	\eeq
	is a so-called non-Hermitian Hamiltonian. The superoperator ${\cal L}$ in \erf{eqn:ME} is the generator of the completely positive 
	map $\cal M$, that is
	\beq \label{maprho}
	\rho(t) = {\cal M}(t) \rho(0) = e^{{\cal L} t} \rho(0).
	\eeq
	For the case of a single jump operator $\hat{c}$ 
	the map can be evaluated as ${\cal M}(t) = \sum_{n=0}^\infty {\cal M}^{(n)}(t)$ where 
	\bqa   \label{map}
	{\cal M}^{(n)}(t) & = & \int_0^t dt_n \int_0^{t_n} dt_{n-1} \cdots \int_0^{t_2} dt_1 {\cal Q}(t-t_n) {\cal J} \times \nonumber \\
	& &  {\cal Q}(t_n-t_{n-1}) {\cal J} \cdots {\cal Q}(t_2-t_1) {\cal J} {\cal Q}(t_1), 
	\eqa
	where ${\cal Q}(t) = e^{t{\cal S}[\hat H']}$ is the propagator of ${\cal S}$. This mapping indicates that, there exists, in principle, a way to monitor the bath such that the conditional state of the system undergoes periods of smooth evolution, represented by $\cal S$, which is interrupted by instantaneous jumps, described by ${\cal J} \equiv {\cal J}[\hat c]$.  
	
	Note that this separation into smooth non-unitary evolution and jumps is not unique. 
	Rather, the master equation \erf{eqn:ME} is invariant under the transformation 
	\bqa
	\hat{c}_l & \rightarrow & \hat{c}_m = \sum_{l=1}^L \widetilde{U}_{ml} \hat{c}_l + \beta_m,  \label{Trans1} \\  
	\hat{H} & \rightarrow &\hat{H} - \frac{i}{2} \sum_{m=1}^M (\beta_m^\ast \hat{c}_m - \beta_m \hat{c}_m\dg), \label{Trans2}
	\eqa
	where $\beta_m$ is a c-number and $\widetilde{U}_{ml} \in {\mathbb C}^{M\times L}$ is an arbitrary semi-unitary matrix. 
	In the context of quantum optics, where the bath is bosonic,  this mathematical 
	transformation has a physical meaning. It corresponds to coherent mixing of the system outputs prior to detection (via $\widetilde{U}$) and adding a weak local oscillator (WLO) of amplitude $\beta_l$ to the $l$-th output field channel prior to detection. 
	Although they give the same master equation, different choices 
	of the parameters {$\{\beta_l\}, \{M_{lk}\}$} give rise to different types of conditional evolution, different unravellings according to \erfa{Trans1}{Trans2}.
	In principle, a sufficiently skilled experimenter could build an apparatus 
	to monitor the bath in a way corresponding to any of the unravellings above. In practice, some unravellings
	may be possible with present experimental techniques and others not at all possible. This will have important 
	implications for which physically realizable ensembles are practically realizable. 
	
	We now impose the further condition of ergodicity. This \blk ensures that in
	the long-time limit the system converges to a unique steady state  satisfying ${\cal L} \rho_{ \rm ss} = 0$. 
	If $\rho_{\rm ss}$ is of finite rank which is always the case for a finite-dimensional system, the stationary state can be expressed 
	in terms of statistical mixture of pure states of a finite set $\{ (p_k, \ket {\psi_k}) \}_{k = 1}^{K}$ with $p_k$ being the probability of 
	occupying $k$th state. Since these states need not be orthogonal in general, there is no upper bound on the number of ways 
	one could do such decompositions. However, it is important to distinguish carefully between ensembles 
	that are  physically realizable (PREs) \blk and those that  are not. \blk The concept which rules out physically unrealizable ensembles
	is referred to as the {\em preferred ensemble fact} \cite{WisVac01}.   The PREs are those that result from some 
	unravelling of the master equation.
	It turns out  that, in general, adaptive monitoring of the environment is necessary to realize a given PRE. That is, 
	it will be necessary to control the above parameters $\{\beta_{m}\}, \{\widetilde{U}_{ml}\}$ based on prior results (jumps). 
	
	One of the essential outcomes of \crf{KarWis11},  which the present paper leans on, is that, 
	for a two-dimensional ergodic quantum system, there is always a two-state PRE, 
	which we will call a TSE for brevity. That is, 
	one can keep track of a qubit using a single-bit memory. We 
	note that the explicit derivation presented 
	in Ref.~\cite{KarWis11} for a qubit does not require a restriction to 
	cyclic jumps, although that is not mentioned.  That is, the system may undergo a `jump' (in the sense of the application 
	of one of the superoperators ${\cal J}[\hat c_l]$) which leaves it in the same state $\ket {\psi_k}$. \blk 
	
	The TSE solution(s) in \crf{KarWis11} may be expressed most easily by rewriting \erf{eqn:MEL} in  the Bloch representation.  It becomes 
	\beq  \label{MEbloch}
	\dot{\vec{r}} = A \vec{r} + \vec{b},
	\eeq
	where $A$ is a $3 \times 3$ 
	matrix, $b$ is a vector with three entries, and $\vec{r}=\an{(\s{x},\s{y},\s{z})\tp}$ is a Bloch vector whose steady state  
	satisfies $A\vec{r}_{ \rm ss} = - \vec{b}$. This steady state is unique if and only if the real part of all eigenvalues of $A$ are negative. \blk In this 
	notation  the TSE is given by $\{ (p_k, \vec{r}_k) \}_{k = \pm}$ where \cite{KarWis11}
	\bqa 
	\vec{r}_k & = & \vec{r}_{ \rm ss} - k \blk \alpha_k \vec{e}_v, \label{eqn:PREs}\\
	p_k & = & \frac{1}{2} \sq{ 1 -  \frac{k \blk{\vec{r}_{ \rm ss}}\cdot{ \vec{e}_v}}{ \sqrt{1-\| {\vec{r}_{ \rm ss}} \|^2 + \ro{{\vec{r}_{ \rm ss}}\cdot{ \vec{e}_v}}^2 }} }\label{eqn:PREp}.
	\eqa
	Here, $\vec{e}_{v}$ is a unit-norm \blk real eigenvector of $A$,  and
	$\alpha_k = k \blk {{\vec{r}_{ \rm ss}}\cdot{ \vec{e}_v}} + \sqrt{1-\| {\vec{r}_{ \rm ss}} \|^2 + \ro{{\vec{r}_{ \rm ss}}\cdot{ \vec{e}_v}}^2 }$. Note that because the Bloch vectors $\vec{r}_+$ and $\vec{r}_-$ have to be real, only those eigenvectors of $A$ that are real yield to the solution. \blk Any 
	real matrix $A$ is  guaranteed to have at least one real eigenvector, 
	and may have more than one. \blk 
	

	\section{State stabilization of a qubit using feedback control} \label{sec:III}
	{In the preceding section we summarized how, conditioned on information obtained by monitoring the environment, 
		the state of the system differs from the solution of the master equation (\ref{eqn:MEL}). Nevertheless, 
		on average, the evolution is unchanged from that master equation solution. However, there is a way 
		to make the information obtained from monitoring affect the average state: by using feedback~\cite{WisMil93c}.} 
	A typical objective of quantum feedback is to stabilize the system into a particular pure state (or close to 
	pure state)~\cite{WisMil94a,HorKil97,WisManWan02,ZhaRusKor05,BusZol06,VijSid12, Bra10, EmaGou14, Ema15}. Recently, a novel way 
	to achieve this in a double quantum dot (DQD) system was proposed~\cite{PolEmaBra11}.
	A DQD is an example of nonequilibrium quantum 
	transport, with two coupled quantum dots connected to leads \cite{NazBla09}. 
	For the analysis in \crf{PolEmaBra11}, \frf{fig:dqd}, two assumptions were made
	to simplify the problem. Firstly, a strong
	Coulomb blockade regime guarantees that at any instant of time there would be {\em at most} one extra electron
	in the system. Secondly, in the high-bias limit {($\mu_L \rightarrow \infty$, $\mu_R \rightarrow -\infty$)\cite{StoNaz96, Bra05}, the 
	resonant energy levels $\epsilon_L$ and $\epsilon_R$ lie within the window between the left and the right lead chemical potential so that the                     
	quantum transport becomes unidirectional, i.e., electrons can only enter the DQD from left and leave it to right.
	The former would imply that the state space of the physical 
	system is spanned by three states. The first is the so-called 
	null state $\ket \emptyset \equiv \ket{N_L , N_R}$ where $N_d$ is 
	 the base number of electrons in dot $d \in \{R, L\}$. The other two are called 
	 single-electron states:
	$\ket L \equiv \ket{N_L+1 , N_R}$, and $\ket R \equiv \ket{N_L , N_R+1}$, with an additional transport electron in the corresponding dot. Under these conditions the system undergoes the transitions $\ket \emptyset \rightarrow \ket L$ and 
		$\ket R \rightarrow \ket  \emptyset$ with tunneling rates $\gamma_L$ and $\gamma_R$, respectively. 
	
	\blk We can describe the evolution of the system using a Lindblad-form ME
	as expressed in \erf{eqn:MEL}; see Ref.~\cite{WisMil10, NazBla09}. The system Hamiltonian is
	\beq \label{sysH}
	\hat{H} = \frac{1}{2} \Delta \hat{\sigma}_z + T_c \hat{\sigma}_x, 
	\eeq
	where $\Delta \equiv \epsilon_L - \epsilon_R$ is detuning, $T_c$ 
	is the coupling constant between the two quantum dots, $\hat{\sigma}_z = |L\rangle \langle L| - |R\rangle \langle R|$, and $\hat{\sigma}_x = |L\rangle \langle R| + |R\rangle \langle L|$. In this work for the sake of simplicity we make the assumption that $\Delta=0$. This means that an electron can be transferred between the two dots at no cost.
	\blk The two Lindblad operators are 
	\beq \label{jumpOp}
	\hat{c}_1 = \sqrt{\gamma_L} \ket L \bra \emptyset,  \qquad \hat{c}_2 = \sqrt{\gamma_R} \ket \emptyset \bra R.
	\eeq
	
	\begin{figure}
		\includegraphics[scale=0.5]{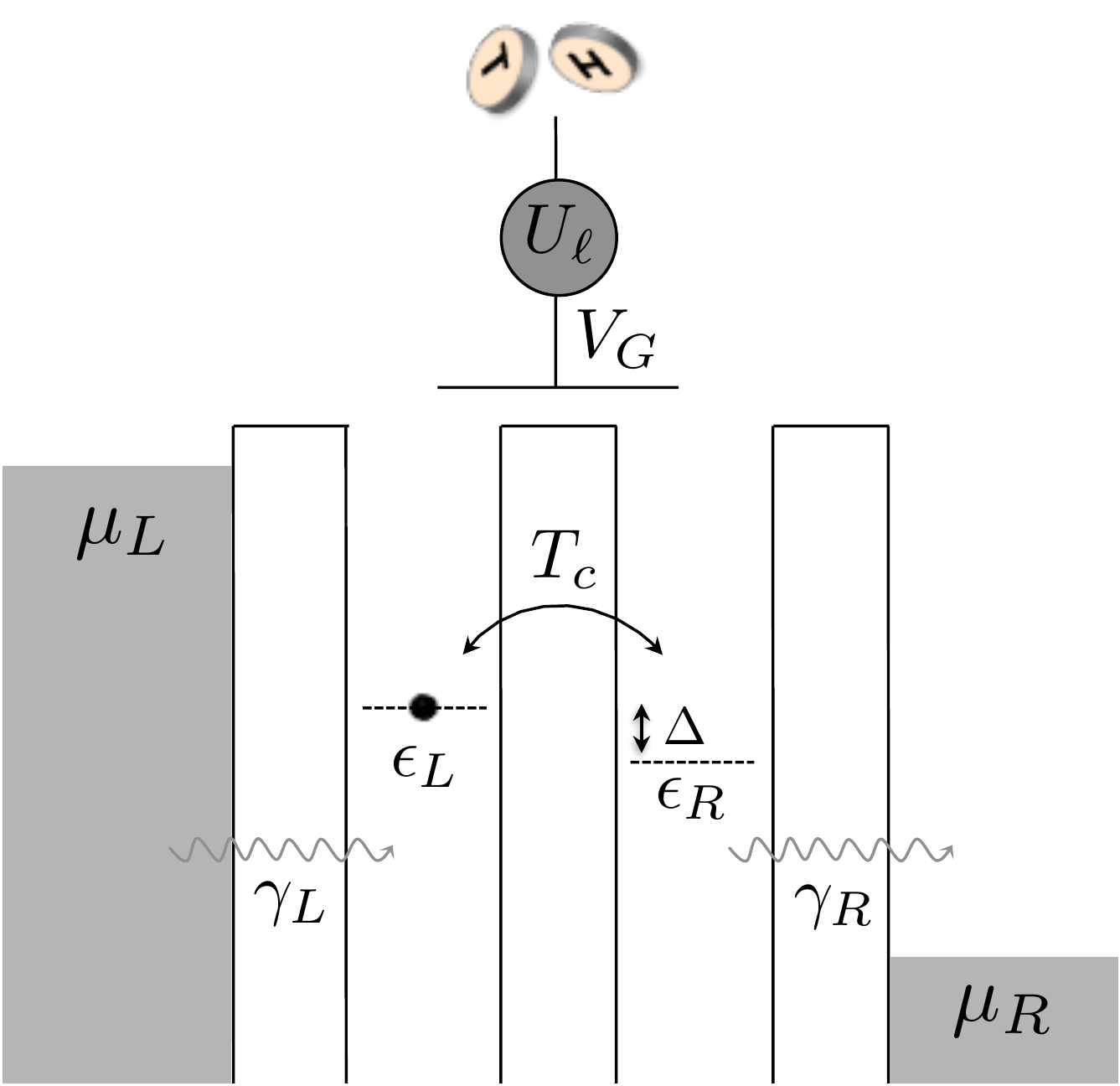}
		\caption{\label{fig:dqd}  Stochastic \blk feedback control of a double quantum dot. Flow of one excess electron is 
			monitored when it tunnels into the left dot at rate $\gamma_L$ (e.g. by a QPC, not shown).
			A unitary control operation (which can be implemented for example with gate voltages $V_G$) then rotates the state 
			of the system into one of two (randomly chosen) desired states. 
			Here $T_c$ is the coupling strength between the dots, $\Delta$ 
			is the detuning, $\gamma_R$ is the tunneling rate out of the right dot, and $\mu_{L(R)}$ are chemical potentials of
			the leads.}
	\end{figure}
	
	Substituting \erfa{sysH}{jumpOp} back into the non-Hermitian Hamiltonian \erf{nonHH} it is straightforward to show that
	in the eigenbasis it takes the following form   
        	\beq  \label{nonHHsys}
        	\hat{H}^\prime =  -i \frac{\gamma_L}{2} \ket {\emptyset} \bra {\emptyset} + \sum_{\ell=\pm} \varepsilon_\ell \ket {\phi_\ell} \bra     
        	 {\widetilde{\phi_\ell}},
        	\eeq
	where
        	\bqa
        	\ket {\phi_\ell} & = & \frac{(i\gamma_R + \ell \kappa)\ket L +4T_c \ket R}{\sqrt{16T_c^2 + |i\gamma_R + \ell \kappa|^2}}, 
	\label{dqd:tse} \\[6pt]
        	\varepsilon_{\ell} & = & -\frac{1}{4} (i \gamma_R + \ell \kappa), \quad \kappa = \sqrt{16 T_c^2 - \gamma_R^2}. \label{def:kappa}
        	\eqa
	Note that the right eigenvectors $\ket {\phi_\ell}$ and the left eigenvectors $\bra {\widetilde{\phi_\ell}}$ of $\hat{H}^\prime$ must be distingusihed 
	from each other since they are not adjoint of each other $\bra {\widetilde{\phi_\ell}} \neq \bra {\phi_\ell}$. 
	Also, the null state has been separated from the other states of the non-Hermitian Hamiltonian due to the orthogonality condition where $\ip 
	{\emptyset}{\phi_\ell} = 0$. Note that the other states are not orthogonal $\ip {\phi_\ell}{\phi_{-\ell}} \neq 0$.

	In Ref.~\cite{PolEmaBra11}, the authors proposed modifying the above dynamics by introducing feedback control, 
	as we know sketch. It is assumed that there is a measurement device, such as a quantum point contact (QPC) \cite{Ubb11,Li13}, 
	in the vicinity of the junction of the left dot to its lead, which can detect single-electron 
	tunneling through this junction. This allows access to the full counting statistics (FCS) \cite{Mar10,Fuj06}
	of single electron quantum jumps.  It is also assumed that the experimenter has the ability to feedback 
	the signal obtained from the integrated detector \blk very fast (effectively instantaneously) to change the state of the system. 
	Immediately after a tunneling event, the state will be $\ket L$, as the following equation shows:   
	\beq  \label{tunL}
	{\cal J} [\hat{c}_1] \ket \emptyset \bra \emptyset  = {\gamma_L} \ket L \bra L.
	\eeq
	The aim of the control is to rotate (in Hilbert space) 
	this into the desired state vector of $\hat{H}^\prime$, that is, $\ket {\phi_\ell}$ \cite{PolEmaBra11}.  
	The control operations can be experimentally implemented e.g., with the gate voltages $V_G$, \frf{fig:dqd}. That is, it is assumed that the 
	following unitary operation can be performed effectively instantaneously: 
	\beq
	\hat U_\ell \ket L = \ket{\phi_\ell}.
	\eeq
	An explicit expression for the unitary operator is given in \arf{appnA}.
	Note that the state $\varrho_\ell = \ket {\phi_\ell} \bra {\phi_\ell}$ is the
	eigenoperator of the smooth evolution superoperator: 
	\beq \label{eigenop}
	{\cal S}\big[\hat{H}^\prime \big] \varrho_\ell = 2 \Im(\varepsilon_\ell) \varrho_\ell.
	\eeq
	
	The system remains in this state as long as the no-jump dynamics continues. However, it will 
	eventually jump back to the null state due to the electron tunneling to the right reservior: 
	\beq \label{tunR}
	{\cal J} [\hat{c}_2] \varrho_\ell= \gamma_R^{\ell} \ket \emptyset \bra \emptyset,
	\eeq
	where $\gamma_R^{\ell} \equiv - 2 \Im(\varepsilon_\ell)$. 
	In \frf{fig:evolution} we summerize this process in the form of a schematic diagram showing the 
	evolution of the conditional state of the system under the control operation. 
	
	The evolution of the system can be described in the Hilbert 
	space ${\mathscr H}^3 \equiv {\rm span} (\ket {\emptyset},\ket {\phi_-}, \ket {\phi_+})$. 
	This indicates that for a given fixed $\ell$ ($+$ or $-$) the system is equivalent to a single resonant level with tunneling 
	rates $\gamma_L$ and $\gamma_R^{\ell}$ as shown in \erfa{tunL}{tunR}. \blk
	The stationary state of the system is 
	\beq  \label{eqn:ss}
	\rho_{\rm ss} = \frac{1}{2\gamma_L+\gamma_R^{\ell}} \big(\gamma_R^{\ell} \ket \emptyset \bra \emptyset + 2\gamma_L \varrho_\ell \big).
	\eeq

	From this, and \erf{def:kappa}, it is apparent that under the condition  $\gamma_L \gg \gamma_R , T_c$, the steady-state probability of 
	finding the system in the null state becomes 
	negligible~\cite{PolEmaBra11}. From the viewpoint of the dynamics, in \frf{fig:evolution}, the time the system spends in the state $\ket \emptyset \bra 
	\emptyset $ in between $\rho_\ell$ and $\ket{L}\bra{L}$ is negligible compared to the time it stays as $\rho_\ell$. (So is, as we have already 
	assumed, the time it takes to perform the operation $U_\ell$.)
	This implies that there is always one transport electron in the system and its state is simply given by $\varrho_\ell = \ket{\phi_\ell}\bra{\phi_\ell}$, depending on which $U_\ell$ was last chosen. This $\ket{\phi_\ell}$ is a pure superposition (\ref{dqd:tse}) of the single-electron states defined above. 

	\begin{figure}
		\includegraphics[scale=0.61]{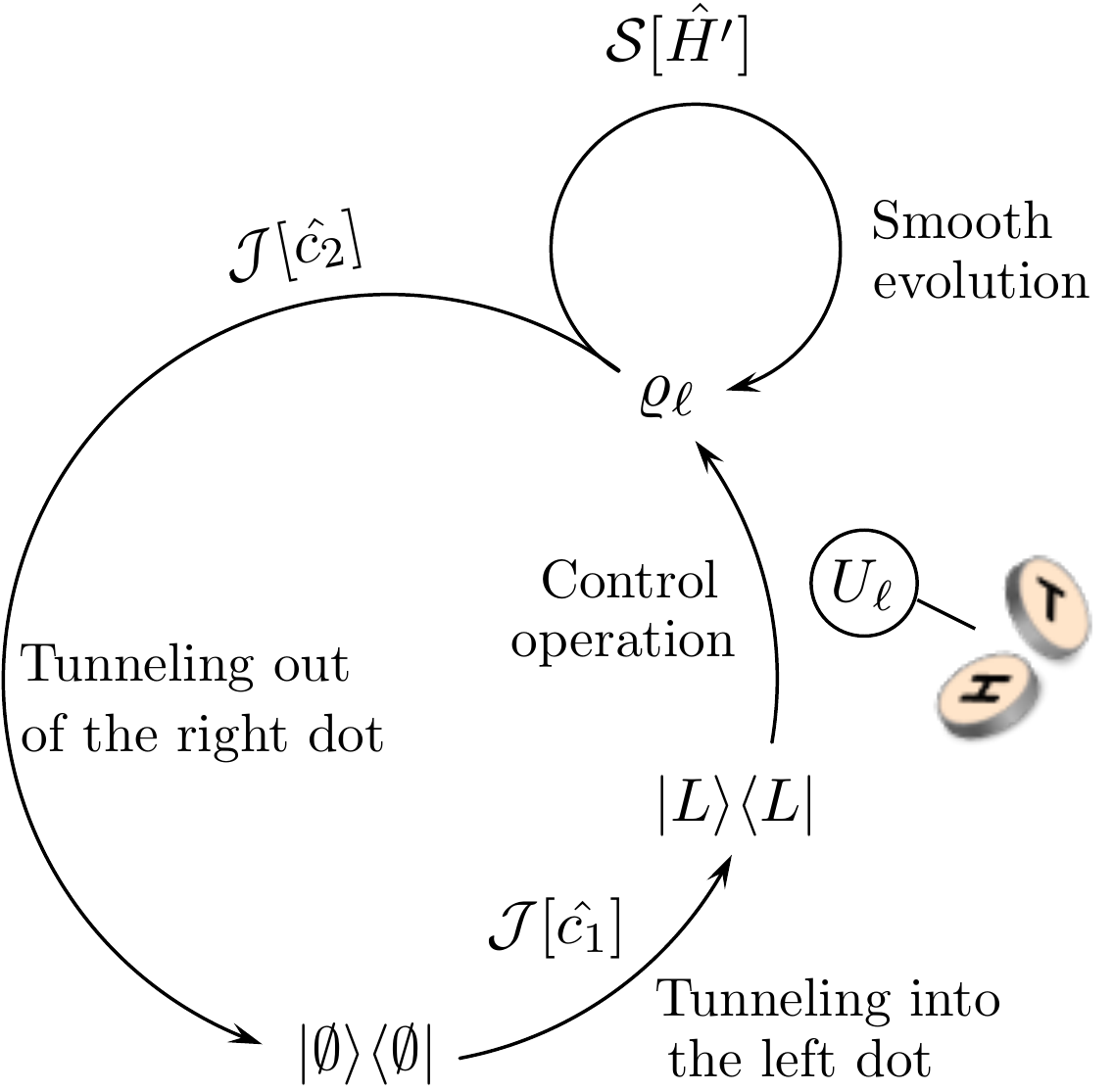}
		\caption{\label{fig:evolution}  A schematic representation of the state evolution of the DQD.}
	\end{figure}

	
	\section{Stochastic Feedback control of the DQD}  \label{sec:IV}  
	Implementing feedback control effectively changes the left jump operator: 
	\beq 
	\hat{c}_1 \to  \hat{c}_1' = \hat{U}_\ell \, \hat{c}_1 = \sqrt{\gamma_L} \, \ket{\phi_\ell}\bra{\emptyset} 
	\eeq
	This gives rise to modification of the ME to 
		\beq  \label{eqn:MEqt}
	\dot{\rho} = -i [\hat H,\rho] + \big( {\cal D} [\hat{c}_1'] + {\cal D} [\hat{c}_{2}] \big) \rho.\eeq
	State purification is achieved in the limit where the tunneling rate into the system is much 
	greater than the tunneling rate out of it, as already mentioned. 
	In this limit we can project out the null state, reducing our description of the system from $D=3$ Hilbert space to that of a qubit. In other words, in 
	the limit where the null state decays much faster than the qubit system, the state $\varrho$ of which can be spanned by $\{ \ket L, \ket R \}$, one can adiabatically 
	eliminate the null state. This means that \erf{eqn:MEqt} reduces to the following (see \arf{appnA} for details) 
	\beq \label{eqn:MEqb1}
	\dot{\varrho} = -i [\hat H, \varrho] + {\cal D} [\hat{c}_{\ell}] \varrho,
	\eeq
where 
	\beq \label{ctrljump}
	\hat{c}_{\ell} = \sqrt{\gamma_R} \, \hat{U}_{\ell} \ket L \bra R = \sqrt{\gamma_R} \, \ket{\phi_\ell} \bra R. 
	\eeq
In this case $\ket{\phi_\ell}$ is still the eigenstate of the non-Hermitian Hamiltonian
\beq 
\hat{H}^\prime = \hat{H} - \frac{i}{2} \hat{c}_\ell\dg \hat{c}_\ell = \varepsilon_\ell \ket {\phi_\ell} \bra {\widetilde{\phi_\ell}}.
\eeq
From this it is apparent that the stationary state is given by $\varrho_\ell$. 
	
	The two possible states $\ket{\phi_\pm}$ are obvious candidates for generating an interesting  
	(nonorthogonal) TSE.  
        However, to apply the theory presented in \srf{sec:II}, one would need 
	a master equation with a non-trivial (rank $>1$) mixture as its steady state. Hence, our intention is 
	not to purify the system into merely one of these 
	states. However, we must not alternate between the two as this would imply non-Markovianity of the system: it requires the memory 
	of the last chosen state for processing the next step. The solution, 
 as shown in Fig.~\ref{fig:evolution}, is that prior to every control operation we toss a coin (figuratively), to determine which operation $\hat U_\pm$ is applied to \blk the   
	system. 
	This means that the system is memoryless and should have a description 
	in terms of a Markovian master equation.  
	If $0 \le \wp_{\ell} \le 1$   is 
	the probability of choosing $\hat U_\ell$, with $\wp_- + \wp_+ = 1$, 
	the master equation is \blk 
	\bqa 
	\dot{\varrho} & = & -i [\hat H,\varrho] + \sum_{\ell = \pm} \wp_{\ell} \, {\cal D} [\hat{c}_{\ell}] \varrho, \label{eqn:MEqb} 
	\eqa 
	and has the solution 
	\beq  \label{stoch_rhoss}
	\varrho_{\rm ss} = \sum_{\ell=\pm} p_\ell \varrho_\ell. 
	\eeq
	Note that the probabilities $p_\ell$ 
	are not, in general, the same as the $\wp_\ell$ in \erf{eqn:MEqb}. We will determine $p_\ell$ when we consider two different 
	regimes of tunneling below.  \blk
	
	 From \erf{eqn:MEqb} one can work out 
	and find the equations of motion in the Bloch representation as in \erf{MEbloch}. Details of matrices $A$ and $b$ are
	given in \arf{appnA}. Therefore, after calculating the real eigenvectors of $A$ 
	and obtaining the steady state
	$\dot{\vec{r}}_{ \rm ss} = 0$ we can plug them in \erfa{eqn:PREs}{eqn:PREp} to construct corresponding
	TSEs. Since the resulting analytical expressions are too intricate to be usefully analyzed we present numerical
	outcomes. Depending on the ratio of the tunneling rate out of the DQD and the coupling constant, there are two different 
	regimes which we call coherent and incoherent in the following. 
	
	
	\begin{figure*} 
		\captionsetup[subfigure]{labelformat=empty}   
		\subfloat[]{\includegraphics[scale=0.43]{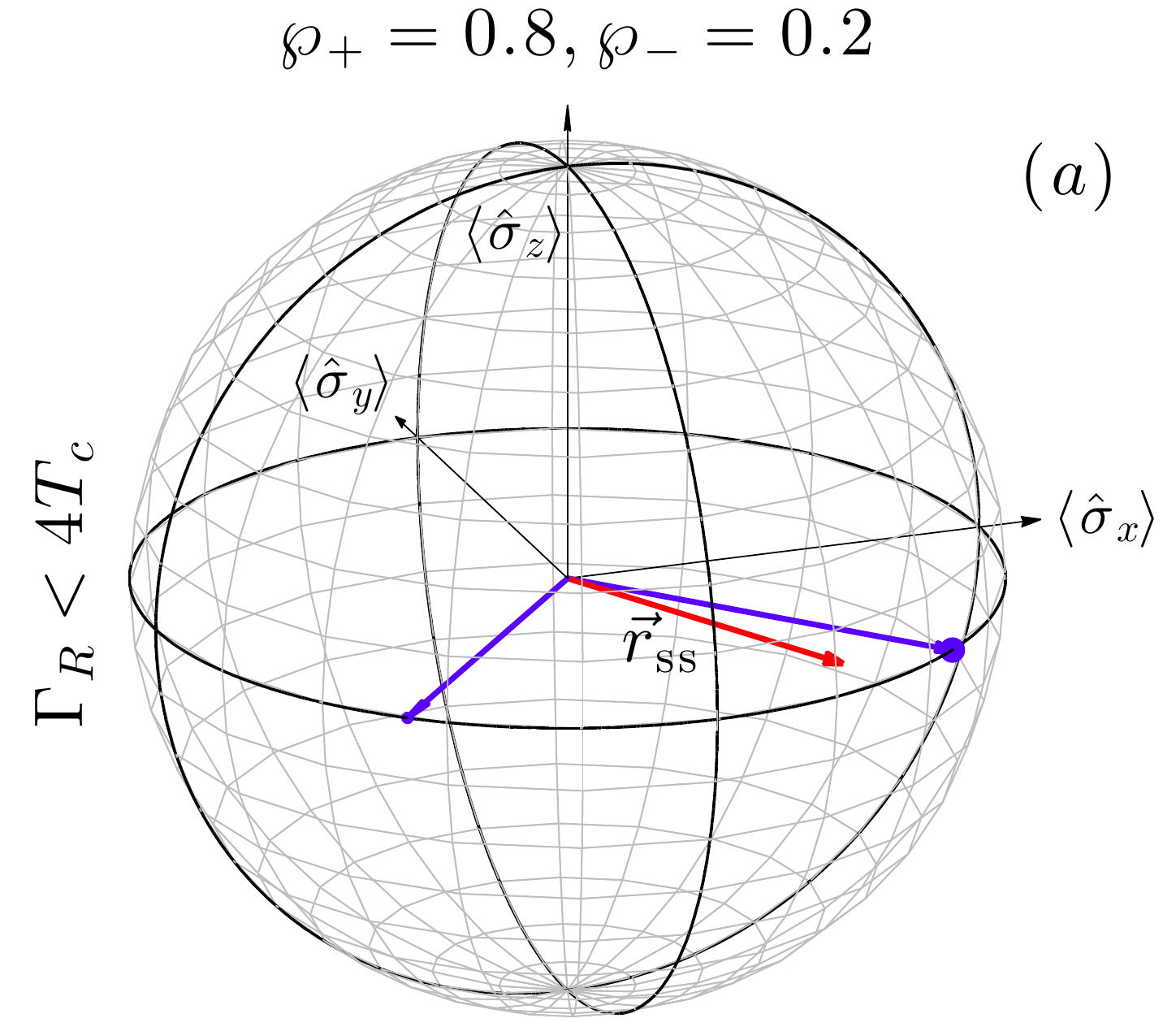} \label{fig:2a} } \,
		\subfloat[]{\includegraphics[scale=0.43]{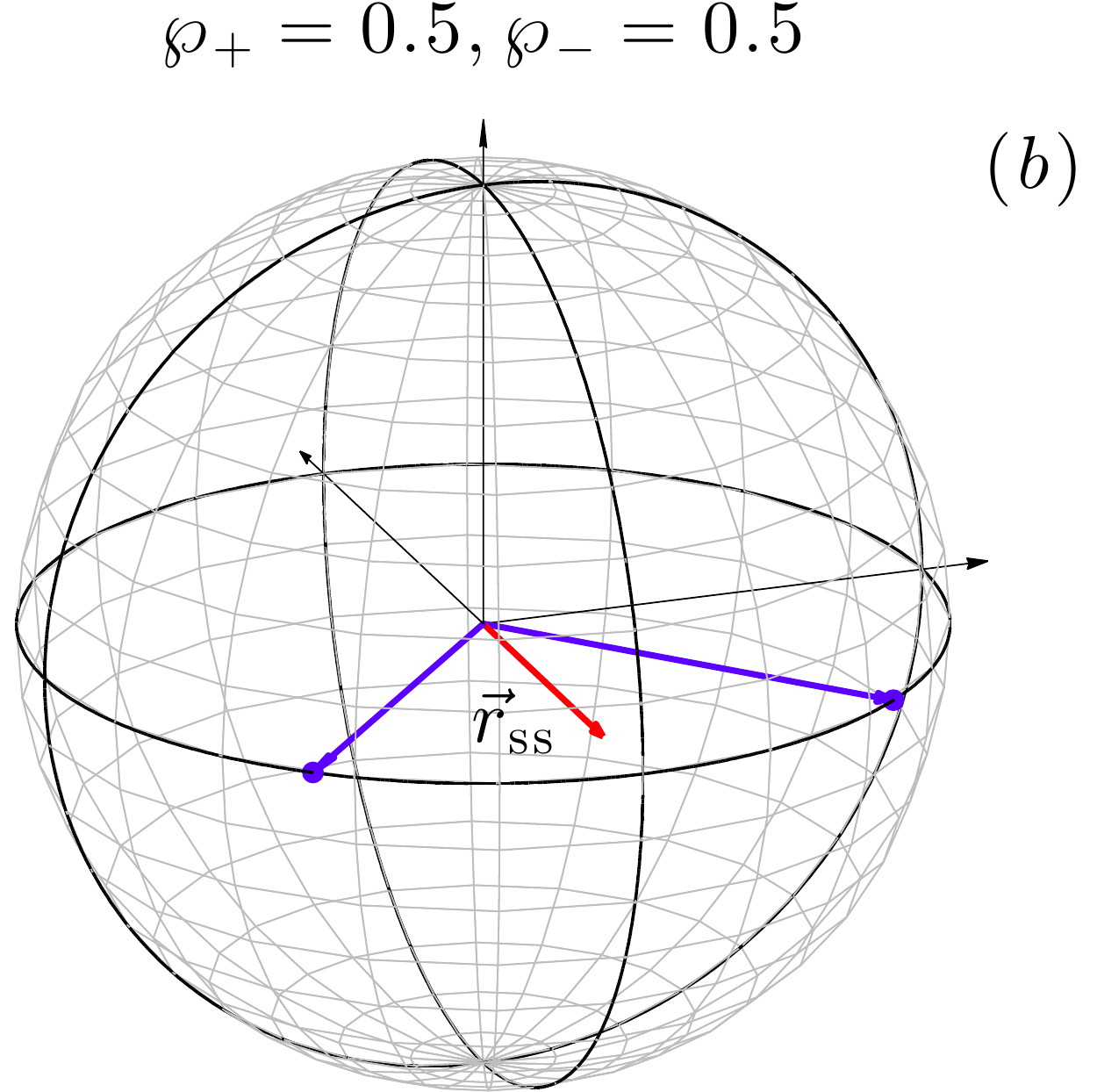} \label{fig:2b}} \, 
		\subfloat[]{\includegraphics[scale=0.43]{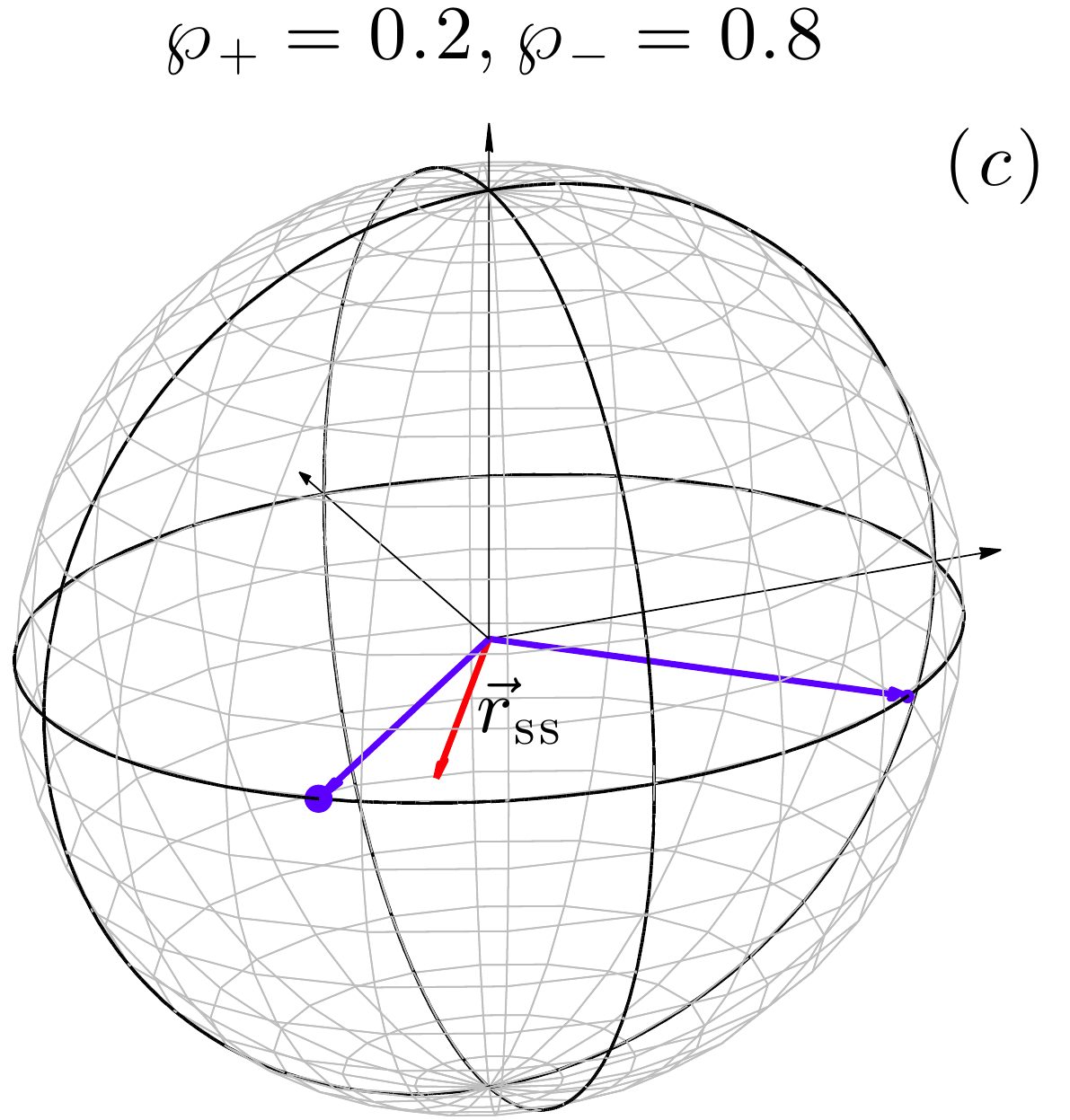} \label{fig:2c} } \\
		\subfloat[]{\includegraphics[scale=0.46]{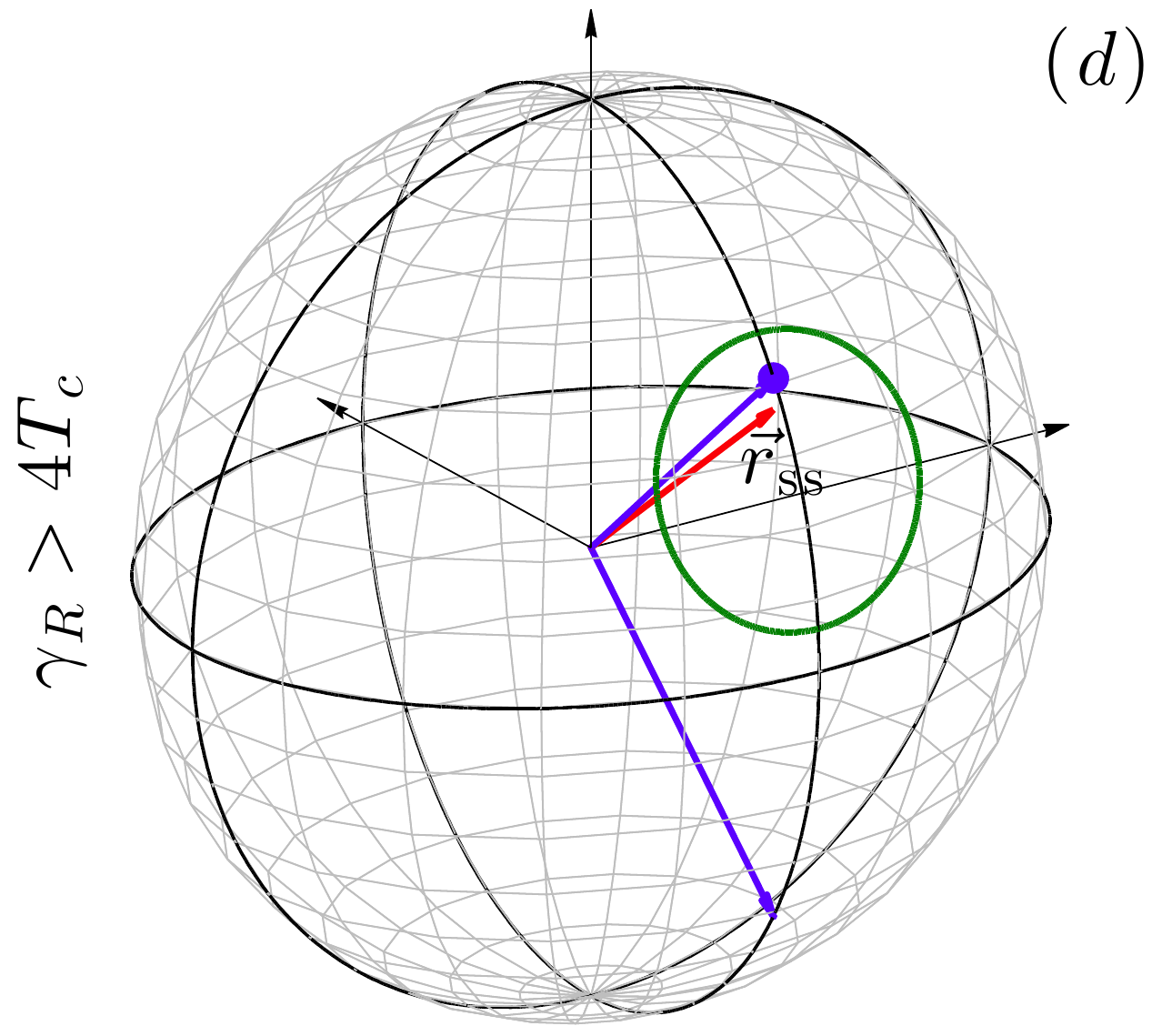} \label{fig:2d} } \;
		\subfloat[]{\includegraphics[scale=0.46]{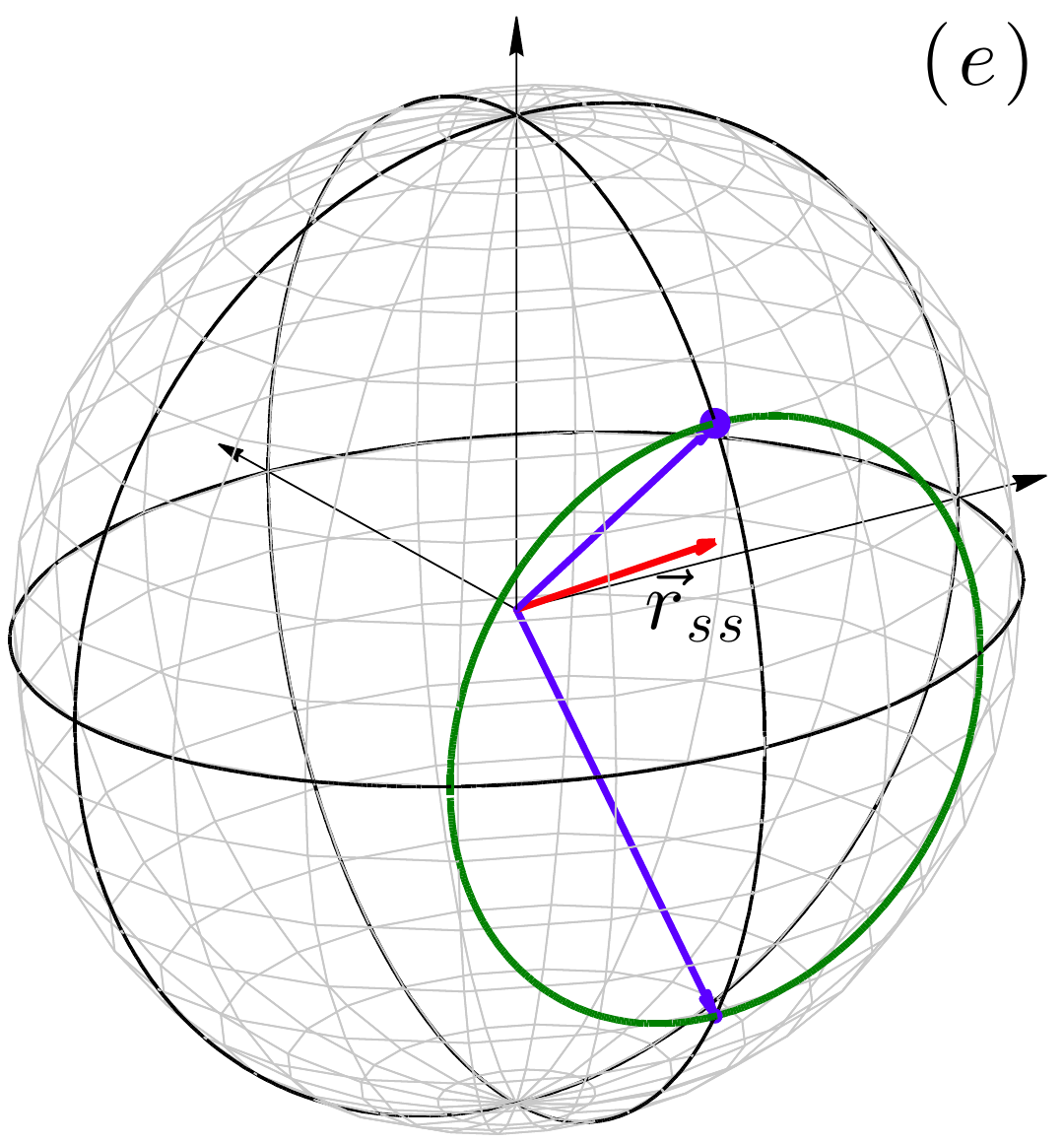} \label{fig:2e}} \;\;\;
		\subfloat[]{\includegraphics[scale=0.47]{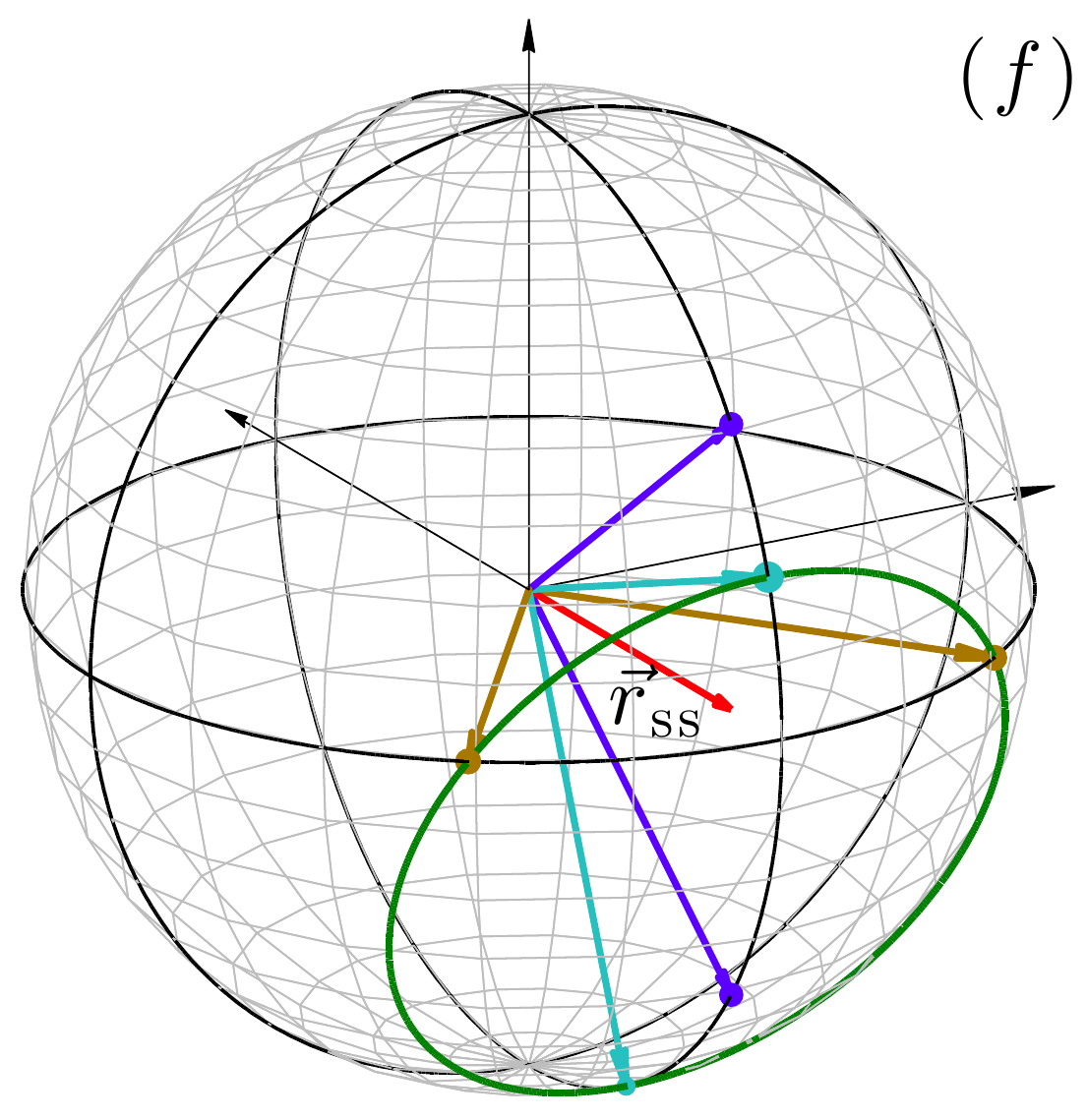} \label{fig:2f} }
		\caption{\label{fig:epsilon=0} (Color online). Bloch representation of TSEs. The top row shows, for $\gamma_R =3, T_c =1$, Bloch vectors of two-state hopping for three different probabilities defining the stochastic feedback: (a) $\wp_+ = 0.8, \wp_- = 0.2$, (b) $\wp_+ = 0.5, \wp_- = 0.5$, and (c) $\wp_+ = 0.2, \wp_- = 0.8$. Blue arrows (dark arrows in the azimuthal plane) represent the state vectors of the TSE with associated probabilities $p_\ell$\, \erf{eqn:PREp}, indicated by  the  volume of the sphere at the tip of each vector. The steady state $\vec{r}_{ \rm ss}$ is depicted by a red arrow. In this regime, the probabilities $\wp_{\ell}$ are equal to the probabilities $p_\ell$ that the system is found in a particular state of the TSE. The bottom row shows TSEs for $\gamma_R = 5, T_c =1$ for the same control probabilities as in the top row. Here, in contrast, in addition to the eigenvectors of the effective Hamiltonian $\hat{H}^{\prime}$, shown by blue arrows (dark arrows in the polar plane), an infinite number of other two-state TSEs exist. The solid green circle depicts the locus of them. By comparing this circle in (d), (e), and (f) it is clear that increasing $\wp_+$ shrinks its radius such that at $\wp_+ = 1$ it eventually collapses to a single point. In (f) we plot the states of two other TSEs each corresponding to a real eigenvector of $A$ shown in light brown (in the azimuthal plane) and cyan (dark arrows in the polar plane). In the bottom row, $\wp_{\ell}$ and $p_\ell$ are entirely different because the latter is a function of the lifetimes of states in the right dot, \erf{pkincoh}. }
	\end{figure*}

 \subsection{Coherent tunneling $( \gamma_R < 4 T_c )$}
	 For the coherent tunneling regime 
	 the steady state is oriented in the equatorial plane. This indicates that the system
	relaxes to a coherent superposition of single-electron states $|L\rangle$ and $|R\rangle$ where the transport electron 
	occupies both quantum dots with some probabilities. The loci of the two states of the TSE also lie in the same plane.
	For the trivial case where $\wp_- = 0$ and $\wp_+=1$ or {\em vice versa,} the steady state reduces to $\vec{r}_{+}$ 
	and $\vec{r}_{-}$, respectively, 
	the two eigenvectors of $H^{\prime}$ which can be represented as Bloch vectors by
	\beq  \label{eqn:BVls}
	\vec{r}_{\ell} = \bigg( - \ell \blk \frac{\kappa}{4 T_c}, -\frac{\gamma_R}{4 T_c}, 0 \bigg),  \;\;\;\; \gamma_R < 4 T_c.
	\eeq
         The top row of \frf{fig:epsilon=0} illustrates TSEs for different purification probabilities for this regime. 
         Since matrix $A$ has only one real eigenvector, therefore there is one TSE.  
         A proportion $p_\ell$ of time that the system spends in a particular state $\rho_\ell$  
	is shown by a little sphere at the end of each arrow. The red 
	arrow shows the locus of the steady state that varies with values of $\wp_{\ell}$. It is noticeable that no matter what 
	values these probabilities are assigned to, as long as $\sum_{\ell} \wp_{\ell} = 1$, they are equal to the corresponding
	$p_\ell$. 
	In order to 
	understand why this is the case we consider the lifetime of states in the right quantum 
	dot given by 
	\beq
	{\tau}_\ell = (\gamma_R \, \varrho_{\ell}^{RR})^{-1},
	\eeq
	where 
	\beq
	\varrho^{RR}_\ell  =  \langle R |\varrho| R \rangle_\ell = \frac{1-\an{\hat{\sigma}_z}_\ell}{2}.
	\eeq
	Since the states in \erf{eqn:BVls} are on the equator we have $\an{\hat{\sigma}_z}_{\pm}=0$, and this further would result in an equal lifetime 
	$\tau_{\ell} = 2/\gamma_R$ for being in the right quantum dot for either of these two states. Hence, the choice of probability $\wp_\ell$ for creating state $
	\varrho_\ell$ simply determines the probability $p_l$ of the system being in the state:
	\beq \label{pkcoh}
	p_\ell = \wp_{\ell}.
	\eeq
	
	\subsection{Incoherent tunneling $( \gamma_R > 4 T_c )$} \label{sec:inctun}
	For the transport electron tunneling incoherently more
	elaborate analysis is required as there are some subtleties. First of 
	all $\tilde{\kappa} = i \kappa$ replaces $\kappa$ in \erf{def:kappa} from which one can easily 
	show that the Bloch vectors corresponding to \erf{dqd:tse} become
	\beq  \label{eqn:BVgt}
	\vec{r}_{\ell} = \bigg(0, -\frac{4 T_c}{\gamma_R}, - \ell \frac{\tilde{\kappa}}{\gamma_R}\bigg),  \;\;\;\; \gamma_R > 4 T_c.
	\eeq
	In contrast to the coherent tunneling regime, it turns out that, in addition to the TSE defined by \erf{eqn:BVgt}, there are an infinite number of other TSEs.  
	As we pointed out at the end of \srf{sec:II}, for any real eigenvector of the 
	matrix $A$ in \erf{MEbloch} there is one TSE. In the regime being considered here, there are at least three such 
	eigenvectors so that each contributes towards    
	formulating a TSE. In addition to the real eigenvector giving the states in  \erf{eqn:BVgt}, 
	there are two other, degenerate, real eigenvectors of $A$.
	  A real linear combination of these other two eigenvectors is also 
	a real eigenvector of $A$. Hence, one can assign a TSE to any such linear combination.  In \frf{fig:2f} 
	two such TSEs are shown in light brown (in the azimuthal plane) and cyan (light arrows in the polar plane), respectively. Since there is a 
	one-parameter continuum of   such linear combinations,  here is a one-parameter family of TSEs. 
	This forms the green circle on all the balls in the bottom row of \frf{fig:epsilon=0}. Note that these ensembles are not necessarily diametrically opposite points on the circle, nor are they equally weighted in general.  Independent of which TSE is being considered, the 
	weighted average of the two unit Bloch vectors is the same steady state Bloch vector. 
For the special case of $\wp_\ell = \wp_{-\ell} = 0.5$, all three eigenvectors are degenerate. In this 
case, the stabilized states of \erf{eqn:BVgt} lie in the green circle as shown in \frf{fig:2e}. 
We note that the stabilized states of \erf{eqn:BVgt}, shown in blue arrows, are strictly independent of $\wp$, while it is evident that 
the locus of pure states on the ball from which the green circle is made up of, does depend on $\wp$. 

Another important point is that no longer are $\{p_\ell\}$ equal to $\{\wp_{\ell}$\}. This is because
	 the states in \erf{eqn:BVgt} would have different lifetimes in the right quantum 
	dot $\tau_\ell = 2/[\gamma_R + \ell \tilde{\kappa}]$. That is, in contrast to the coherent tunneling regime 
	the probability $p_\ell$ is dependent upon $\tau_\ell$ and can be expressed as
	\beq \label{pkincoh}
	p_{\ell} = \frac{\wp_{\ell} {\tau}_{\ell}}{\sum_{j=\pm} {\wp_j {\tau}_j}},
	\eeq
	where the probability for preparing the $\ell$-th state $\wp_{\ell}$ is weighted by the relevant lifetime of the states in the right quantum dot. This is clearly evident from \frfs{fig:2e}{fig:2f}. Note that should the lifetimes were equal $\tau_1=\tau_2$, this relation would reduce to \erf{pkcoh} as expected.
	
	At the critical parameter value, $\gamma_R = 4 T_c$, the steady state is a pure state
	that the system ultimately relaxes into $\vec{r}_{ \rm ss} = \vec{r}_{\pm} = (0, -1, 0)$. It can be imagined that 
	as one approaches this setting from either side, for a fixed chosen values of $\wp_\ell$'s, the Bloch vectors shown in blue in \frf{fig:epsilon=0}, start to get 
	closer to each other and ultimately they become the same up to the point that the green circle shrinks to a point. 
	
	\subsection*{Physical meaning of the other TSEs}   \label{subsec:IIIA}
	We have already seen that there is a measurement and feedback control scheme to realize the ME given in \erf{eqn:MEqb} such that the conditional states of the system are described by \erf{eqn:BVgt}. This raises the question: if this protocol does not realize the other infinite number of TSEs, then in what sense are they physically realizable? 
It may be \blk that the hypothetical ensembles are in principle physically realizable but through the instantiation of the aforementioned ME, \erf{eqn:MEqb}, they are not practically realizable. If one were to
induce the evolution described by this ME in a completely different way, then the other
TSEs might be realized in practice. In what follows we present our speculation on the possibility of 
realizing these ensembles in the system as described. 

It is necessary to consider a grand system composed of two subsystems: the DQD and the whole measurement and feedback control process (MFCP) with state vectors $\ket{\phi_{\ell}}$ and $\ket{\xi}$, respectively, in the spirit of autonomous (passive) feedback control \cite{EmaGou14, Ema15}. One would have to imagine that the state of the grand 
system is in a pure entangled form as the coherent  superposition of these state vectors 
\beq  \label{Gstate}
\ket{\Psi} \propto \sum_{r} {\ket{\phi_{\ell | r}}} {\ket{\xi_r}}.
\eeq
Here the measurement record $r$ determines the state of the DQD. This implies that the entire MFCP itself is being treated as a quantum system which could be monitored. However, an important point is that if this monitoring reveals the record $r$ used in the feedback loop, 
then the system would simply collapse into one of the two states $\ket{\phi_\pm}$ given in \erf{eqn:BVgt}.  
It is conceivable that it would be possible to obtain the other ensembles in Sec.~\ref{sec:inctun} by employing a quantum 
eraser method  \cite{KimScu00, WalMon02} for the whole MFCP.
However, when it comes to consider the fermionic systems, the superselection rules (SSR) also should be taken into account as additional constraints \cite{BarSpe07}. For instance, the transformation in \erf{Trans1} does not hold for fermionic systems; fermionic fields cannot have a non-zero mean value, 
unlike bosonic fields. Thus, we tend towards a negative opinion on the possibility of realising these other ensembles of Sec.~\ref{sec:inctun} in the controlled quantum dot system.

	\section{Counting Statistics of quantum transport}   \label{sec:V}
	Just as the mean current through the DQD gives information about the dynamics of the 
	system, its fluctuation, variance and maybe higher order cumulants provide yet further
	information. One can determine these from the
	probability distribution, $P(n,t) = {\rm Tr}[\varrho^{(n)}(t)]$. Here $P(n,t)$ is the probability that $n$ electrons tunnel into 
	the right lead in time $t$ (assuming that transport is unidirectional from the source to the drain), and $\varrho^{(n)}(t)$ represents
	the system state matrix at time $t$ with $n$ quantum jumps from which electrons have been transferred to the drain.
	This process of electron counting may be realized experimentally {\em e.g.} by using a QPC as a 
	detection apparatus. 	It is easy to show that using the microscopic field-detector theory for a state with $n$ excitations
	the {\em n}--resolved form of the ME 
	given in \erf{eqn:ME}, where $\varrho (t) = \sum_{n=0}^\infty \varrho^{(n)}(t)$, for the DQD with feedback control is the following 
	\cite{Mar10}  
	\beq \label{eqn:nME}
	\dot{\varrho}^{(n)} =  {\cal S}\varrho^{(n)} + \sum_{\ell=\pm} \wp_\ell {\cal J}[\hat{c}_\ell] \varrho^{(n-1)}. 
	\eeq
	Since this equation can be solved in $n$--space only for a finite $n$ which subsequently requires diagonalization of a 
	tridiagonal matrix, it is thus more convenient to use the Fourier transform $\dot{\varrho}(\chi) = \sum_n e^{i n \chi} \varrho^{(n)} $ with 
	the {\em counting field} $\chi$ as the conjugate variable of $n$. Then \erf{eqn:nME} in $\chi$--space 
	becomes
	\beq \label{eqn:chiME}
	\dot{\varrho}{(\chi)} =  \big({\cal S} + \sum_{\ell=\pm} \wp_\ell {\cal J}[\hat{c}_\ell] e^{i \chi}\big) \varrho{(\chi)} \equiv {\cal L}(\chi) \varrho(\chi). 
	\eeq
	The cumulant generating function ${\cal F}(\chi,t)$ is defined as \cite{Mar10}
	\beq 
	e^{{\cal F}(\chi,t)} = \sum_{N=0}^{\infty} P(n,t) e^{i n \chi },
	\eeq
	where by the definitions of the Fourier transform and the probability distribution mentioned above
	it is transformed into
	\beq \label{eqn:CGF}
	{\cal F}(\chi,t) = {\rm ln} \big( {\rm Tr \big[ e^{{\cal L}(\chi)(t-t_0)} \varrho(t_0) \big]}\big).
	\eeq
	Since in the long time limit $t\rightarrow \infty$ the system is essentially in the steady state, it is thus
	possible to use the eigendecomposition of ${\cal L}(\chi)$ to further simplify \erf{eqn:CGF} to 
	\beq \label{eqn:CGFsimplified}
	{\cal F}(\chi,t) = \lambda_0(\chi) t + {\rm constant},
	\eeq
	where $\lambda_0$ is the largest eigenvalue of the spectrum of ${\cal L}(\chi)$. Therefore, 
	the $n$th-order zero frequency current correlation is given by
	\bqa \label{eqn:curcor}
	\an {S^{(n)}} & = & \frac{d}{dt} \frac{\partial^n}{\partial(i \chi)^n} {\cal F}(\chi, t) |_{\chi=0, t\rightarrow \infty} \nn \\
	& = & \frac{\partial^n}{\partial(i \chi)^n} \lambda_0 (\chi) |_{\chi=0}.
	\eqa
	The first order current correlation $\an {S^{(1)}}$ is just the mean current. Thus it would be more convenient to
	normalize the higher order current correlation functions to this to obtain the zero-frequency Fano factor \cite{Mar10}
	\beq \label{eqn:FF}
	F^{(n)} = \frac{\an {S^{(n)}}}{\an {S^{(1)}}}.
	\eeq
	
	In \frf{fig:fcs} we plot the Fano factor up to order $n=4$ as a function of $\gamma_R/T_c$ to study both coherent and 
	incoherent tunneling regimes. For each regime a different set of control parameters ($\theta, \vartheta_c$) are 
	chosen, see \arf{appnA} for details. It is evident that assigning different values to the probabilities $\wp_{\pm}$ in 
	\erf{eqn:chiME} does not affect the statistics of the current when the system is in the coherent tunneling 
	case, \blk $\gamma_R \le 4T_c$. In fact, the Fano factor remains unchanged at $F^{(n)} = 1$ 
	revealing the Poissonian behavior of the system (left panel of \frf{fig:fcs}). However, as $\gamma_R$ 
	exceeds $4T_c$, that is in incoherent tunneling regime, Fano factors start to deviate from Poisson 
	statistics (right panel of \frf{fig:fcs}). Again, this behavior can be
	rooted in the fact that the lifetimes of the two states are different, \erf{pkincoh}. 
	Note that for the trivial case where the feedback control is not stochastic, that is, $\wp_+ = 1$ and $\wp_- = 0$, or {\em vice versa}, all 
	Fano factors become unity. As expected, this is in complete agreement with the results of 
	\crf{PolEmaBra11} where the system is merely stabilized into one of the two feasible states.   
	
	\begin{figure}
		\includegraphics[scale=0.88]{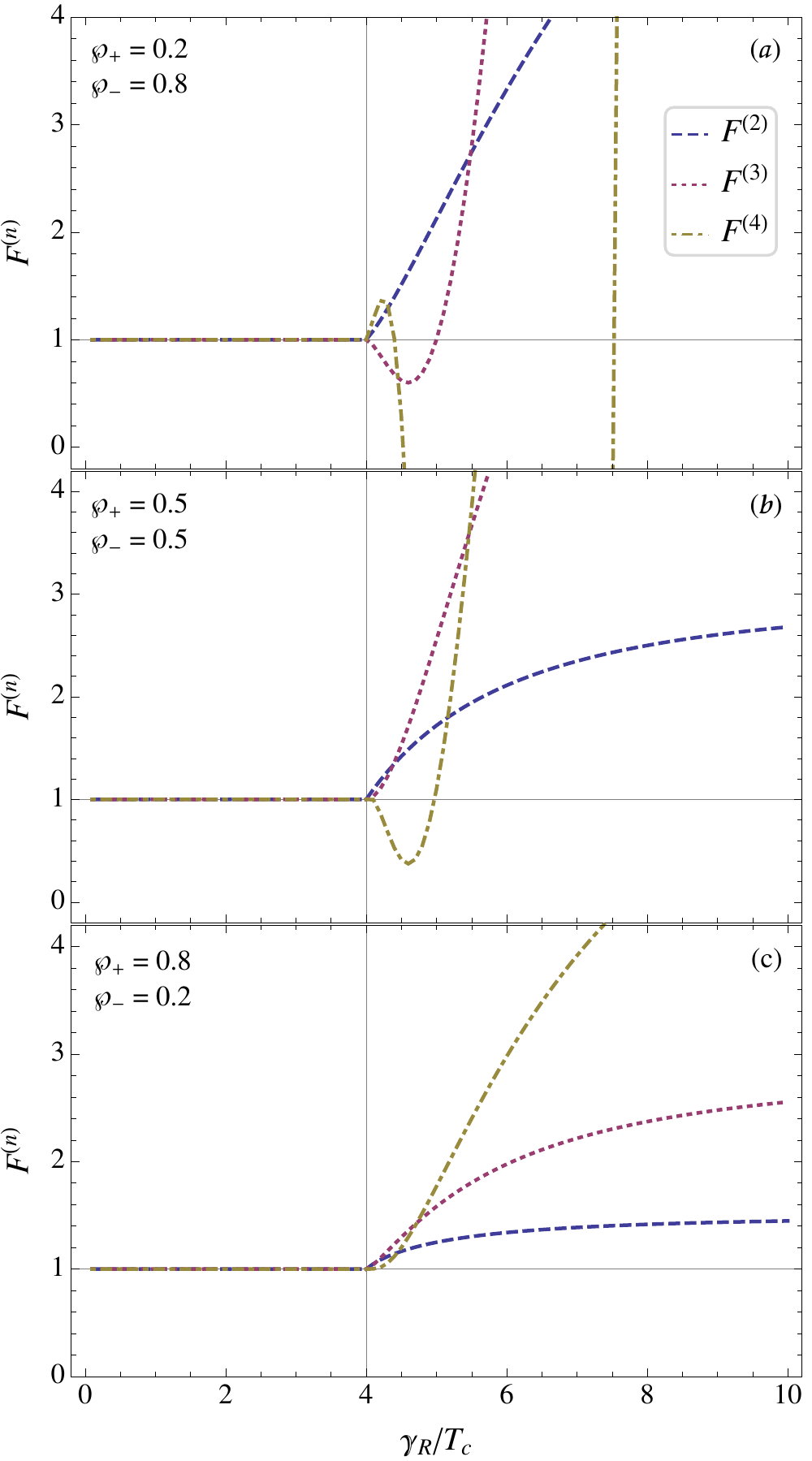}
		\caption{\label{fig:fcs} (Color online). The value of Fano factors $F^{(n)}$, \erf{eqn:FF}, for the DQD versus the ratio of 
			tunneling rates $\gamma_R/T_c$: $n = 2$ (blue dashed), $n = 3$ (pink dotted), and $n = 4$ (light brown dashed-dotted).
			Three different sets of probabilities are shown (a) $\wp_+ = 0.2, \wp_- = 0.8$, (b) $\wp_+ = 0.5, \wp_- = 0.5$, and (c) 
			$\wp_+ = 0.8, \wp_- = 0.2$. 
			As long as $\gamma_R \le 4T_c$
			all Fano factors are equal to one no matter what value is assigned to $\wp_{\pm}$. This is correspond to the Poissonian statistics
			of the current which is independent of the order of the current correlation. \blk
			In contrast, as $\gamma_R$ becomes larger than $4T_c$ Fano factors depend 
			on the probabilities $\wp_{\pm}$ to the extent that $F^{(n)} = 1$ only holds if and only if $\wp_+ = 1$ and $\wp_- = 0$, or {\em vice versa}.}
	\end{figure}

\section{Conclusion}
In summary, we have shown that using a particular stochastic feedback control scheme 
for a double quantum dot, 
it is possible to engineer Markovian open quantum system dynamics such that the conditioned state of the single-electron qubit is (essentially) always one of two non-orthogonal pure states. Here `conditioned' means conditioned on the measurement signal derived from the tunneling into or out of the system and used 
in the control loop. Previous schemes to realize conditionally such a non-orthogonal finite ensemble of pure states \cite{KarWis11,DarWis14} relied upon an adaptive measurement. Adaptive measurement is a distinct type of control from feedback control in that the unconditioned state of the system is unaffected by the control in the former case, but is affected in the latter. 
In the context of quantum optics, a suitable  adaptive 
measurement can be performed by controlling the amplitude of a weak 
local oscillator to the system output field \cite{CooGer07, YonFur12}.

Because of the complexity of performing adaptive measurements, and the inefficiency of photon collection and detection, no 
such adaptive measurement experiment to realise a non-orthogonal pure-state ensemble has been realized. 
In contrast, for the solid-state feedback scheme we have considered here, one conditions on 
 the detections that are natural for the system. 
Another factor that may motivate one to implement this scheme is the access to a sensitive 
electrometer such as a QPC which can be highly efficient in detecting quantum jumps \cite{ChuChe15}.
Thus the system we have studied is a promising candidate for practically approximating an 
interesting (non-orthogonal) highly-pure TSE.

The double dot system has two regimes, depending on the strength of the inter-dot tunneling 
to the rate of tunneling out of the system. In the incoherent (weak)  
tunneling regime system has dynamics distinct compared to the case of coherent (strong) tunneling regime. 
This transition corresponds to an exceptional point \cite{Hei04} of the non-Hermitian Hamiltonian $\hat{H}^\prime$, eq. (13), and thus bears some analogy with simple models of quantum bifurcations or phase transitions. Physically, this is a manifestation of the spectral line effect discovered by Dicke in 1953 \cite{Dic53}. 
The combination of feedback with a stochastic choice of the control operation provides a tool to probe this transition via current fluctuations:
First, the full counting statistics analysis demonstrates 
that as long as the control operation chooses randomly between the states of the TSE, the system statistics 
shows non-Poissonian behavior. Secondly, due to the fact the two nonorthogonal states have different lifetimes, the 
occupation probabilities $p_\ell$ are not set purely by the probabilities $\wp_\ell$ in the stochastic control.
 And finally, infinitely many two-element non-orthogonal pure-state ensembles exist that the system could, in 
 principle, be purified to. However, the feedback control scheme just realizes one. That is, even though all of 
 them are in principle physically realizable only one is practically realizable by the feedback loop. 

 There remains some future work to be done 
for addressing some of the practical issues that would happen in a real experiment. These may 
include efficiency, time-delays in the feedback loop, how fast the control operation could be 
performed, how to implement the control unitary and what imperfections could arise in this. 
From the fundamental side, there is an open question to be explored which addresses the last point 
mentioned above: is there any physical way to realize the ensembles that are theoretically possible 
but which do not arise from the feedback loop we construct? This issue might be connected to an 
autonomous formulation of the stochastic feedback loop, including a better understanding of the 
thermodynamics of this form of control \cite{StrEsp13}.
	
	\section*{Acknowledgement}

This research was supported by the ARC Centre of Excellence Grant No. CE110001027 and DFG SFB 910. 
	
	\appendix 
	\numberwithin{equation}{section}
	\section{Supplemental Material}  \label{appnA} 
	\subsection{Adiabatic elimination of the null state}
	 In the limit in which the damping rate of the null state is much more quicker than the decay rate of the 
	 qubit system, the null state can be projected out by employing  
	  an adiabatic elimination technique~\cite{WarWis00}. In the following we concisely describe how this is done. 
	The state of the qutrit system can be expressed as
	\beq \label{rho3d}
	\rho = \varrho + \delta \op{\emptyset}{\emptyset}
	\eeq
	where $\varrho$ is spanned by $\{ \ket L, \ket R\}$ in the qubit subspace and $\delta = O (\gamma_R/\gamma_L)$. It is easy to show 
	that the evolution of $\delta$ is obtained according to
	\bqa 
	\dot{\delta} & = & \bra \emptyset \dot{\rho} \ket{\emptyset} \nonumber \\
	 & = & -\gamma_L \delta + \gamma_R \bra R {\varrho} \ket{R},
	\eqa
	where we have used \erf{eqn:MEqt}. In the limit of $\gamma_L \gg \gamma_R$ where the decay rate of the 
	null state is much faster than the rate at which the qubit system damps, one can employ the adiabatic elimination which yields to
	\beq \label{delta}
	\delta \approx \frac{\gamma_R}{\gamma_L} \bra R {\varrho} \ket{R}.
	\eeq
	The dynamics of the qubit state is also obtained by
	\beq  \label{varrhodot}
	\dot{\varrho} = \hat{\Pi}_{\rm qb} \, \dot{\rho} \, \hat{\Pi}_{\rm qb}
	\eeq
	with $\hat{\Pi}_{\rm qb} = \op{L}{L} + \op{R}{R}$ being the projector to the qubit subspace. Thus, using \erfa{eqn:MEqt}{delta} it is straightforward to 
	show that \erf{varrhodot} is transformed into
	\bqa
	\dot{\varrho} & = & - i [H, \varrho] + \gamma_R \op{\phi_\ell}{\phi_\ell} \bra{R} \varrho \ket{R} - \nonumber \\
	& & \frac{\gamma_R}{2} \big( \op{R}{R} \varrho + \varrho \op{R}{R} \big) \nonumber \\
	& = & - i [H, \varrho] + {\cal D}[\hat{c}_\ell] \varrho.
	\eqa

         \subsection{Bloch equations and control parameters}
         \blk
	Given the ME in \erf{eqn:MEqb} matrices $A$ and $b$ of the equation of motion read 
	\bqa  \label{eqn:A12}
	A_{11} & = & -2 \gamma_R \sum_{\ell = \pm} \wp_{\ell} \big( |u_y^{\ell}|^2 + |u_z^{\ell}|^2 \big),   \\
	A_{12} & = & \gamma_R  \sum_{\ell = \pm} \wp_{\ell} \Big[ \big( u_x^{\ell} {u_y^{\ell}}^*  + {u_x^{\ell}}^* u_y^{\ell} \big) - \nn \\
	&  & i \big( u_z^{\ell} {w^{\ell}}^* - {u_z^{\ell}}^* w^{\ell} \big) \Big]  - \Delta, \\         
	A_{13} & = & \gamma_R  \sum_{\ell = \pm} \wp_{\ell} \Big[ \big( u_x^{\ell} {u_z^{\ell}}^*  + {u_x^{\ell}}^* u_z^{\ell} \big) + \nn \\
	& & i \big( u_y^{\ell} {w^{\ell}}^* - {u_y^{\ell}}^* w^{\ell} \big) \Big], \\
	A_{21} & = & \gamma_R  \sum_{\ell = \pm} \wp_{\ell} \Big[ \big( u_x^{\ell} {u_y^{\ell}}^*  + {u_x^{\ell}}^* u_y^{\ell} \big) + \nn \\
	&  & i \big( u_z^{\ell} {w^{\ell}}^* - {u_z^{\ell}}^* w^{\ell} \big) \Big]  + \Delta, \\
	A_{22} & = &  -2 \gamma_R \sum_{\ell = \pm} \wp_{\ell} \big( |u_x^{\ell}|^2 + |u_z^{\ell}|^2 \big),  \\
	A_{23} & = &  \gamma_R  \sum_{\ell = \pm} \wp_{\ell} \big( u_y^{\ell} {u_z^{\ell}}^*  + {u_y^{\ell}}^* u_z^{\ell} \big) - \nn \\
	& &  i \big( u_x^{\ell} {w^{\ell}}^* - {u_x^{\ell}}^* w^{\ell} \big) \Big] - 2T_c,  
	\eqa            
	\bqa  \label{eqn:A39}   
	A_{31} & = &   \gamma_R  \sum_{\ell = \pm} \wp_{\ell} \Big[ \big( u_x^{\ell} {u_z^{\ell}}^*  + {u_x^{\ell}}^* u_z^{\ell} \big) - \nn \\
	& & i \big( u_y^{\ell} {w^{\ell}}^* - {u_y^{\ell}}^* w^{\ell} \big) \Big], \\
	A_{32} & = &  \gamma_R  \sum_{\ell = \pm} \wp_{\ell} \big( u_y^{\ell} {u_z^{\ell}}^*  + {u_y^{\ell}}^* u_z^{\ell} \big) + \nn \\
	& &  i \big( u_x^{\ell} {w^{\ell}}^* - {u_x^{\ell}}^* w^{\ell} \big) \Big] + 2T_c,  \\
	A_{33} & = &  -2 \gamma_R \sum_{\ell = \pm} \wp_{\ell} \big( |u_x^{\ell}|^2 + |u_y^{\ell}|^2 \big).
	\eqa
	and 
	\begin{eqnarray}  \label{eqn:b}
		b_{1} & = & i\, 2 \gamma_R \sum_{\ell = \pm} \wp_{\ell} \big( u_y^{\ell} {u_z^{\ell}}^*  - {u_y^{\ell}}^* u_z^{\ell} \big), \\                                  
		b_{2} & = &  i \,2 \gamma_R \sum_{\ell = \pm} \wp_{\ell} \big( u_z^{\ell} {u_x^{\ell}}^*  - {u_z^{\ell}}^* u_x^{\ell} \big),   \\
		b_{3} & = &  i\, 2 \gamma_R \sum_{\ell = \pm} \wp_{\ell} \big( u_x^{\ell} {u_y^{\ell}}^*  - {u_x^{\ell}}^* u_y^{\ell} \big).
	\end{eqnarray} 
	
	The complex 4-vector $\ut{u}^{\ell} = (u_x^{\ell}, u_y^{\ell}, u_z^{\ell},w^{\ell})$ 
	has the following components
	\begin{eqnarray}  \label{eqn:u}
		u_x^{\ell} & = & \frac{1}{2} \Big[{\rm cos}\big(\vartheta_c^{\ell}/2\big) - i\,{\rm sin}\big(\vartheta_c^{\ell}/2\big) {\rm cos}\big(\theta^{\ell}\big) \Big],   
		\\[4pt]
		u_y^{\ell} & = & i \,u_x^{\ell}, \\[4pt]
		u_z^{\ell} & = &  \frac{i}{2}\, {\rm sin}\big(\vartheta_c^{\ell}/2\big) {\rm sin}\big(\theta^{\ell}\big), \\[4pt]
		w^{\ell} & = & - u_z^{\ell}.
	\end{eqnarray}  
	See below for definition of $\theta^\ell$ and $\vartheta_c^{\ell}$ separately for each tunneling regime. 
	The control operator then can be expressed as a function of these parameters
	\beq \label{ctrlopt}
	\hat{U}_{\ell} = e^{- i \vartheta_c^{\ell} \vec{n}_\ell.\vec{\sigma}/2},
	\eeq
	which rotates a state around the unit 
	vector $\vec{n}_\ell = ({\rm sin}\theta^\ell,0,{\rm cos}\theta^\ell)$ by $\vartheta_c^{\ell}$. 
	Note that $\vartheta_c $ in this paper is as twice as the control parameter $\theta_C$ in \crf{PolEmaBra11}. 
	
	For the coherent tunneling regime where $\gamma_R < 4 T_c$ the control parameters are given by 
	\bqa
	\theta^{\ell} & = & {\rm arccos} \big( \frac{\ell \, \kappa}{\sqrt{32 T_c^2 - \gamma_R^2}} \big), \label{theta_coh} \\
	\vartheta_c^{\ell} & = & 2\,{\rm arccos} \big( \frac{\gamma_R}{\sqrt{32} T_c} \big), \label{thetac_coh}
	\eqa
	and for $\gamma_R > 4 T_c$ where an electron tunnels incoherently are given by
	\bqa
	\theta^{\ell} & = & \pi/{2}, \label{theta_incoh}\\
	\vartheta_c^{\ell} & = & 2\,{\rm arctan} \big( \frac{4 T_c}{( \gamma_R + \ell \,\tilde{\kappa})} \big). \label{thetac_incoh}
	\eqa
	For the case where $\gamma_R = 4 T_c$ they are equal to $\theta^{\ell} = \vartheta_c^{\ell} = \pi/2$.


\end{document}